\documentstyle[prb,aps]{revtex} 
\input epsf
\tighten
\begin{document}
\draft
\twocolumn[\hsize\textwidth\columnwidth\hsize\csname @twocolumnfalse\endcsname
\title{Universality, frustration and conformal invariance\\
in two-dimensional random Ising magnets}
\author{F. D. A. Aar\~ao Reis, S. L. A. de Queiroz,
and Raimundo R. dos Santos}
\address{
Instituto de F\'\i sica, Universidade Federal Fluminense, Avenida Litor\^anea
 s/n, 24210-340 Niter\'oi RJ, Brazil}
\date{\today}
\maketitle
\begin{abstract}
We consider long, finite-width strips of Ising spins with randomly distributed
couplings. Frustration is introduced by allowing both ferro- and 
antiferromagnetic interactions. Free energy and spin-spin correlation 
functions are calculated by transfer-matrix methods. 
Numerical derivatives and finite-size scaling concepts allow estimates of the 
usual critical exponents $\gamma/\nu$, $\alpha/\nu$ and $\nu$ to be obtained,
whenever a second-order transition is present.
Low-temperature ordering persists for suitably small 
concentrations of frustrated bonds, with a transition governed by pure--Ising
exponents. Contrary to the unfrustrated case, subdominant terms
do not fit a simple, logarithmic-enhancement form. 
Our analysis also suggests a vertical critical line at and below the Nishimori 
point. 
Approaching this point along either the temperature axis or the Nishimori
line,
one finds non-diverging specific heats. 
A percolation-like ratio $\gamma/\nu$ is found upon analysis of the
uniform susceptibility at the Nishimori point.
Our data are also consistent with frustration inducing a breakdown of the 
relationship between correlation-length amplitude and critical exponents, 
predicted by conformal invariance for pure systems.    
\end{abstract}

\narrowtext
\vskip2pc]

\section{Introduction}

Frustration induced by quenched randomness may have rather complex effects on
the behaviour of spin systems~\cite{rieger}. The Edwards-Anderson
model~\cite{ea}
for spin glasses is one on which a great deal of theoretical effort has been 
concentrated, mostly on account of the basic simplicity of its formulation.
In the version that will be of direct interest to us, one  has Ising spins 
interacting through couplings of the same strength and random sign 
($\pm J$ Ising spin glass model). For the symmetric case of equal 
concentrations of $+$ and $-$ signs,  a two-dimensional ($d =2$) system
with only first-neighbour interactions is paramagnetic at all
temperatures~\cite{byom,by}. 
Recent results, according to which addition of a suitable set of
second-neighbour couplings would stabilise a low-temperature spin-glass 
phase~\cite{lc}, have been disputed on grounds that the ordering thus
observed is a finite-size effect~\cite{parisi}.

However, for {\em asymmetric} distributions of ferro- and
antiferromagnetic bonds one  can have long-range order in $d=2$ with only
first-neighbour interactions (see Ref.~\onlinecite{domany} for references 
to early work) on a square lattice, for suitably small concentrations of 
frustrated plaquettes~\cite{domany}. 
Experimental work on frustrated magnets 
X$_2$Cu$_x$Co$_{1-x}$F$_4$, X = K~\cite{jamet} or Rb~\cite{schins}, has
shown the existence of an ordered state in such conditions, though direct 
quantitative comparison with theoretical calculations for $\pm J$ Ising 
systems is not straightforward because (1)  ferromagnetic (Cu--Cu)  bonds 
are weaker than antiferromagnetic (Co--Co)  by a factor of $\sim 8$, and 
(2) the magnetism of Cu is Heisenberg-like with a small Ising anisotropy, 
which induces additional complications such as transverse freezing. 
   
Estimates of the location of, and properties along, the critical line of the 
$\pm J$ model have been produced~\cite{ozni,kiog}. Further, since the
pioneering work of Nishimori~\cite{nish} asymmetric Ising spin glasses have 
been shown to display quite unique features, in all space dimensionalities. 
Among these is the so-called {\it Nishimori line} (NL) along which the 
internal energy can be calculated exactly, and which roughly separates the 
regions in the temperature-randomness parameter space where either
ferromagnetic or spin-glass correlations dominate. 
Accordingly, the intersection of the NL with the para-ferromagnetic boundary 
is of special significance, even in $d=2$ where no spin-glass phase is 
expected at non-zero temperature ($T \neq 0$). 
It has been proposed~\cite{ldh} that such intersection, to be referred to as
{\it Nishimori point} (NP), coincides with the multicritical point expected 
to exist along the boundary.
Very recently, the NP has been investigated in $d=2$ and 4 by series 
analysis~\cite{adler}, and in $d=2$ also by a variant of the 
Chalker-Coddington~\cite{cc} network model~\cite{cho}. 
Numerical values of critical exponents thus obtained are very close to those 
of percolation~\cite{stauffer} in $d=2$ (though not in $d=4$~\cite{adler}).   

On the other hand, $d=2$ unfrustrated random Ising systems, 
such as the random-bond (i.e., bonds being $J$ or $rJ$, $0<r<1$),
and the diluted model, have been the subject of renewed interest 
over the past years, for two main reasons.
Firstly, different scenarios regarding the universality class of these
systems have been proposed: {\em weak universality} versus {\em logarithmic
``corrections''.}
In the former, critical exponents are distinct from those in the pure case,
but their ratios to $\nu$, the correlation length exponent, remain the same
as in the pure case~\cite{weak}; 
in the latter, the pure-system power-law critical behaviour is reinforced 
by logarithmic divergences~\cite{dstheory}.
Secondly, the applicability of conformal invariance to random spin systems 
has not been put on grounds as firm as those for pure
systems~\cite{cardy}, thus (for instance) the corresponding
relationship between critical exponents and correlation length 
amplitudes needs to be checked in each case. 

Thus, a systematic study of the asymmetric $\pm J$ Ising model on a square 
lattice is of interest, not only in relation to specific quantitative 
questions (such as the shape of the critical line and its intersection with 
the NP), but also in relation to the broader context of universality classes
in disordered systems, singling out the effects of frustration.
With this in mind, here we will focus on the following main questions:
(i) to  what extent, if any, do logarithmic corrections to pure-system 
behaviour describe criticality for small degrees of frustration?
(ii) does the connection between critical exponents and correlation length 
amplitudes hold in the case?  
(iii) can we provide evidence for (or against) the conjecture~\cite{adler} 
that the critical behaviour at the $d=2$ Nishimori point is percolation-like? 
These issues are addressed through the calculation of free energies and 
spin-spin correlation functions on long, finite-width strips of a square 
lattice.
The rate of decay of correlation functions determines correlation lengths 
along the strip. 
We have already shown how averaged values of such quantities, and their 
numerically calculated field-- and temperature derivatives, enable one to 
extract critical properties of unfrustrated disordered 
models~\cite{sbl,msbl}. 
For the latter class of systems
in particular, we gave numerical evidence in favour of the logarithmic 
corrections scenario; we also showed that the relationship between critical 
exponents and correlation length amplitudes, predicted by conformal
invariance~\cite{cardy}, remains valid provided one uses averaged 
correlation lengths~\cite{sbl,sldq}. 
The validity of conformal invariance ideas  for (unfrustrated)
disordered  $q$-state Potts models has also been 
verified~\cite{cj,berche}.

This paper is organized as follows. In Sec.\ \ref{calc} we outline numerical 
aspects of our calculational procedures, as applied to the asymmetric $\pm J $ 
Ising model.
Results for the phase boundary and critical behaviour above the Nishimori line 
are discussed in Sec.\ \ref{above}, while the Nishimori point and the 
region below it are discussed separately in Sec.\ \ref{NP}.
Our findings are then summarized in Sec.\ \ref{concl}.

\section{Calculational method}
\label{calc}

We have used long strips of a square lattice, of width $4 \leq L \leq 14$
sites with periodic boundary conditions across the strip. 
Only even widths were used, in order to accommodate possibly-occurring 
unfrustrated antiferromagnetic ground states. 
We compute spin-spin correlation functions along the ``infinite'' direction 
by transfer-matrix methods~\cite{nig82,fs1,fs2}, extracting averaged 
correlation lengths. 
By the same methods we numerically obtain the free energy and its second 
derivatives with respect to $(i)$ a uniform external field, which are used in 
connection with  finite-size scaling (FSS) for estimating $\gamma/\nu$; 
and $(ii)$ temperature, again used with FSS concepts for estimating 
$\alpha/\nu$.
In order to provide samples that are sufficiently representative of disorder, 
we iterated the transfer matrix~\cite{fs2} typically along $10^7$ ($10^8$ near
the NP) lattice spacings.

At each step, the respective vertical and horizontal bonds between 
first-neighbour spins $i$ and $j$ were drawn from a probability distribution 
\begin{equation}
 P(J_{ij})= p\ \delta (J_{ij} -J_0) + (1-p)\ \delta (J_{ij} +J_0) \ .
\label{eq:1}
\end{equation}
For a square lattice the phase diagram in the
$T - p$ plane is invariant with respect to the symmetry $p \leftrightarrow 1-p$;
thus we shall restrict ourselves to $0.5 < p \leq 1$, meaning that bulk
antiferromagnetic order will play no part in what follows.

The direct calculation of correlation functions, 
$\langle\sigma_{0} \sigma_{R}\rangle$, 
goes according to Section 1.4 of Ref.~\onlinecite{fs2}, with the corresponding 
adaptations for an inhomogeneous system~\cite{sldq}. 
For fixed distances up to $R=50$, and for strips with lengths as given above, 
the correlation functions are averaged over an ensemble of $10^5$--$10^6$ 
different estimates to yield $\overline{\langle\sigma_{0} \sigma_{R}\rangle}$.
The average correlation length, $\xi^{av}$ (which carries a dependence on $T$,
$p$ and strip width $L$), is in turn defined by
\begin{equation}
\overline{\langle\sigma_{0} \sigma_{R}\rangle}
\sim
\exp\left(-R/\xi^{av}\right),\\
\label{xi}
\end{equation}
and is calculated from least-squares fits of straight lines to semi-log 
plots of the average correlation function as a function of distance,
in the range $10\leq R\leq 50$. 
Finally, $\xi^{av}$ is itself averaged over the different realizations of 
disordered bonds. 

In this context it must be recalled that, although in-sample fluctuations of 
correlation functions do not die out as strip length is increased, averaged 
values converge satisfactorily~\cite{dqrbs}; 
as done before~\cite{sbl}, here we make use of this fact to calculate error 
bars of related quantities.

\begin{figure}
\epsfxsize=8,5cm
\begin{center}
\leavevmode
\epsffile{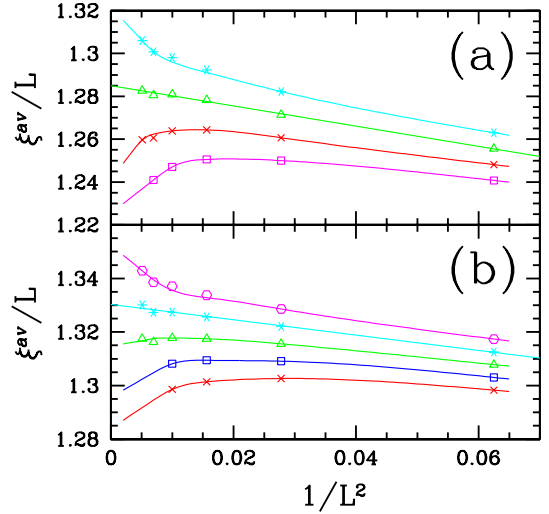}
\caption{Size-scaled correlation length versus $1/L^2$ ($L$ is the strip 
width) at different inverse temperatures, $K\equiv J_0/k_BT$ in the low 
frustration regime: 
(a) $p=0.95$ and $K=0.535$ (stars), 0.534 (triangles), 0.533 (crosses),
and 0.532 (squares);
(b) $p=0.92$ and $K=0.637$ (hexagons), 0.636 (stars), 0.635 (triangles), 
0.634 (squares), and 0.633 (crosses). 
Lines drawn through the points are guides to the eye.
See Table \protect{\ref{table1}} for corresponding estimates of $K_c$.
}
\label{fig:xi/L}
\end{center}
\end{figure}

In contrast with the unfrustrated disordered models considered 
earlier~\cite{sbl,msbl}, here the exact critical temperature is not known as a
function of $p$, so our first step was to use averaged correlation lengths 
together with FSS ideas~\cite{nig82,fs1,fs2} to obtain an approximate
critical curve $T_c(p)$. 
This approach is safe because the only underlying assumption is that a 
second-order phase transition occurs, without further hypotheses on its 
universality class.
In the usual phenomenological renormalisation recipe, used for pure systems, 
one looks for the fixed point $T^{\ast}$ of 
$\xi^{av}(L,T^{\ast},p)/L =
\xi^{av}(L^{\prime},T^{\ast},p)/L^{\prime}$ 
(in the case one would use $L^{\prime} = L-2$). 
For disordered systems it should be stressed that, even if logarithmic 
corrections are present in the bulk limit, the (averaged) correlation length
at the critical point should still scale linearly with the strip width $L$, to 
leading order~\cite{sbl}. 
Thus, here we produce estimates of $T_c(p)$ by scanning a range of temperatures 
for fixed $p$, and bracketing the interval for which $\xi^{av}/L$ appears to 
approach a finite value as $L \to \infty$.
The width of such temperature interval gives the respective error bar, as 
illustrated in Fig.~\ref{fig:xi/L}.
Two remarks are in order in relation to this approach.
Firstly, this is more convenient here than the standard fixed-point search,
since (1) intrinsic uncertainties associated to the individual $\xi^{av}$
are amplified when estimating $T^{\ast}(L,L^{\prime})$, and (2)
it is the extrapolation as $L, L^{\prime} \to \infty$ that
matters in the end.
As the range of available strip widths is not very
broad, it is important that, for given $T$,
the sequence of data from individual $L$'s has one more point,
and also slightly smaller error bars, than that of  
$T^{\ast}(L,L^{\prime})$.
And, secondly, with our procedure one already gains an
insight into corrections to scaling: by varying the power of $1/L$
against which $\xi^{av}/L$ is plotted, one can check how better to
produce (inclined) straight lines within the bracketed temperature range
singled out by the initial scan. It must be stressed that the
location and width of the bracketed
range itself, separating the (high-temperature) regime in which one is certain
that $\xi^{av}/L \to 0$ and that (low-temperature) in which
$\xi^{av}/L$ diverges, are practically insensitive to the choice of power.
Indeed, though in Fig.~\ref{fig:xi/L} we plotted 
$\xi^{av}/L$ {\it vs.} $L^{-2}$ -- inspired by results for pure~\cite{dds}
and unfrustrated random~\cite{sldq} systems --, we have also checked that 
using $\xi^{av}/L$ {\it vs.} $L^{-1}$,  $L/\xi^{av}$ {\it vs.} $L^{-1}$ or
$L^{-2}$, 
changes no significant digits of our extrapolated estimates.

Once, for fixed $p$, one has an estimate of $T_c(p)$, the next step is to
calculate the critical free energy and its appropriate derivatives. This is done
by evaluating the largest Lyapunov exponent $\Lambda_{L}^{0}$
for strips of width $L$ and length  $N \gg 1$~\cite{glaus,ranmat}.
The average free energy per site is 
$f_{L}^{\ av}(T) = - {k_BT \over J_0 L} \Lambda_{L}^{0}$, in units of $J_0$.
The initial susceptibility and specific heat of a strip 
are then given by:
\begin{equation}
\chi_L (T) ={ \partial^{2}  f_{L}^{\ av}(T) \over \partial
h^2}\Biggr|_{h=0}
;\ \ 
C_L (T) ={ \partial^{2}  f_{L}^{\ av}(T) \over \partial
T^2}\Biggr|_{h=0};
\label{eq:chiC}
\end{equation}
where 
$h$ is an external field coupling to the order parameter; the size dependence 
of these quantities will be discussed below.
We shall take $h$ as uniform (ferromagnetic order), which is reasonable 
for low frustration; at the Nishimori point, this choice implies singling out
one of the two scaling directions (more on this below). An extensive discussion
of calculational details is given, for the specific heat, in
Ref.~\onlinecite{msbl}. Here we recall that, since the derivatives are
numerically obtained by calculating {\it e.g.}
$2f_L (T_c) - f_L (T_c + \delta T) - f_L (T_c - \delta T)$ with
$\delta T = 10^{-3}T_c$, sample-to-sample 
fluctuations are roughly as large as the difference between free energies at
these three temperatures; thus one must ensure that {\it 
the same configuration} of bonds (that is, the same sequence of pseudo-random 
numbers) is used in the comparison of different temperatures: free energies
of the same bond geometry have to be subtracted. The probable errors
for the free energy differences are then much smaller than those for the free
energies themselves. Similar procedures were used in a transfer-matrix study of
interface energies in random systems~\cite{derv}. The same argument applies for
the susceptibilities, substituting $\delta h$ (typically of order $10^{-4}$
in units of $J$) for $\delta T$.   

Finally, one should have in mind that the inverse of $\xi^{av}$ is, in 
principle, distinct from the difference between the two leading Lyapunov 
exponents, which gives the decay of the most probable, or {\em typical} 
(as opposed to averaged) correlation 
function.\cite{sbl,sldq,dqrbs,ranmat,lyapcor}
Nonetheless, for $d=2$ unfrustrated disordered Ising systems they have turned 
out to be numerically very close~\cite{sldq,dqrbs}, at least at the
critical 
point (see below for remarks on low-temperature behaviour in the present
case);
significant differences arise only in the corrections to scaling, which are 
relevant for extrapolation to the thermodynamic limit~\cite{lyapcor}.
In Refs.~\onlinecite{ozni,kiog}, the model considered here was studied with the 
aid of typical correlation lengths, $\xi^{typ}$, also calculated on strip 
geometries, but disregarding corrections to scaling; 
below, we will comment on some of the differences between our results and 
theirs.

\begin{table}
\caption{
Inverse critical temperatures for low frustration. The
pure-system
value of $\xi/L$ is $4/\pi = 1.2732 \dots$, from conformal invariance.}
\vskip 0.2cm 
 \halign to \hsize{\hskip 1.6cm\hfil#\hfil&\hfil#\hfil&\hfil#\hfil\cr
  $p$ & $K_{c}(p)$       & $\lim_{L\to\infty} \xi/ L$\cr
 0.99 & $0.4555\pm 0.0005$ &  $1.275\pm 0.015$ \cr
 0.95 & $0.534\pm 0.001$ &  $1.285\pm 0.015$\cr
 0.92 & $0.6360\pm 0.0015$ &  $1.325\pm 0.015$\cr}
 \label{table1}
\end{table}

\section{Above the Nishimori point}
\label{above}

We start by applying the above-described procedure to scale correlation 
lengths, for $p$ close to 1. 
Figure \ref{fig:xi/L} displays $\xi/L$ vs.\ $1/L^2$ at different  
temperatures, for (a) $p=0.95$ and (b) $p=0.92$.
The corresponding estimates for the critical temperatures are shown in 
Table \ref{table1}, and compare rather well with those of
Refs.~\onlinecite{ozni} and \onlinecite{kiog}. 
Using exact and approximate data, respectively at $p=1$ and 0.99, 
the reduced slope of the critical curve at the pure point is estimated to
be $\left. {1 \over T_c(1)}{dT_c\over dp} \right|_{p=1}=3.25\pm 0.11,$
which compares very
well with the exact result, 3.2091~\cite{domany}.

\begin{figure}
\epsfxsize=8,5cm
\begin{center}
\leavevmode
\epsffile{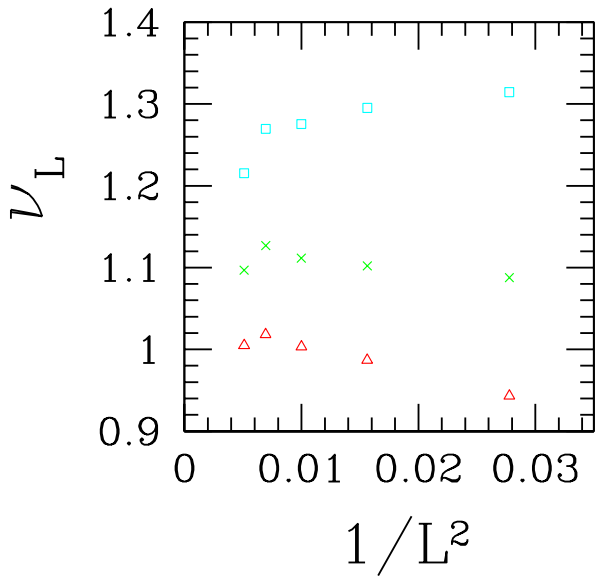}
\caption{Size dependence of $\nu_L$ as given by Eq.\ (\protect{\ref{nul}}), for
$p=0.99$ (triangles), 0.95 (crosses), and 0.92 (squares). Error bars are
smaller than data points.}
\label{fig:nul}
\end{center}
\end{figure}
We now turn to the correlation-length exponent, $\nu$.
Since $\nu$ does not appear explicitly in the expression for 
$\xi_L^{av}(T_c)/L$, one resorts to the temperature derivative of the 
correlation length, which can also be cast in a similar scaling form,\cite{sbl}
\begin{equation}
\mu_L\equiv{d\xi^{av}_L\over dt} =
L^{1+{1\over\nu}}\ {\cal G} (z),\ z\equiv{\xi_\infty(t)\over L},
\label{mul}
\end{equation}
with $t \equiv (T - T_c)/T_c$, and ${\cal G}$ is a finite-size scaling 
function. 
Assuming a simple power-law divergence $\xi_{\infty} \sim t^{-\nu}$, -- i.e., 
ignoring, for the time being, less-divergent terms such as logarithmic 
corrections --  we obtain the estimates for systems of sizes $L$ and $L-2$:
\begin{equation}
{1 \over \nu_L} = 
{ \ln \left(\mu_L/\mu_{L-2}\right)_{T=T_c}
\over \ln (L/L-2)} - 1,
\label{nul}
\end{equation}
where the derivatives are calculated numerically at the extrapolated (i.e., 
$L\to\infty$) value of $T_c$. 
For fixed $L$ and $T_c(p)$, we obtain one estimate of $\nu_L$ for each disorder 
configuration; these estimates are then averaged over different disorder 
configurations to yield the data shown in Fig.\ \ref{fig:nul}, for $L=6 - 
14$.
The trend displayed in Fig.\ \ref{fig:nul} is dramatically different from
the one observed in the case of unfrustrated disorder:\cite{sbl}
All curves (for different values of $p$) show a distinct downturn  
(as $L$ increases), 
and a limiting value $\nu_\infty=1$, common to all values of $p$ considered,
becomes more likely.

This should be contrasted with the case of unfrustrated disorder, for which
no downturn was observed, and the extrapolations indeed seemed to indicate 
a steady convergence towards a disorder dependent exponent 
$\nu=\nu(p)\geq\nu_{\rm pure}=1$; c.f. Ref.\ \onlinecite{sbl}. 
In that case, available theories~\cite{dstheory} pointed towards
describing those apparent exponents as resulting from 
power-law divergences with pure-system exponents, enhanced by
multiplicative logarithmic terms; such
expectations were later confirmed through transfer matrix
calculations on strips~\cite{sbl,msbl}.
Though in the present case the downturn in the trend may be taken as 
indicative that these corrections are absent, considerable insight should
be gained by trying to fit the data along similar lines.

\begin{figure}
\epsfxsize=8,5cm
\begin{center}
\leavevmode
\epsffile{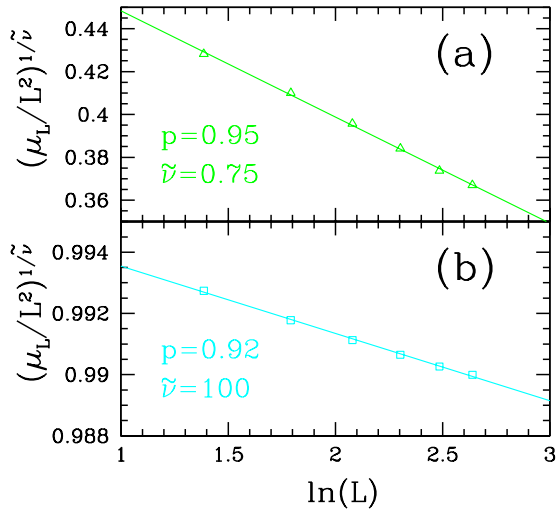}
\caption{Fits of Eq.\ (\protect{\ref{eq:mul2}}) to determine $\tilde\nu$, for
$p=0.95$ (triangles) and 0.92 (squares). 
Error bars are smaller than data points.}
\label{fig:mul}
\end{center}
\end{figure}

The forms of logarithmic corrections in random systems have been derived 
within a field-theoretic approach~\cite{dstheory}, which does not 
explicitly account for frustration effects.
Nonetheless, inspired by our experience with unfrustrated disorder,
we decided to check whether in the present case such
corrections also arise.
The theory~\cite{dstheory} which successfully accounts for
unfrustrated disorder 
predicts that the correlation length of the disordered Ising model, near 
the critical point, is given by
\begin{equation}
\xi_{\infty}\sim t^{-\nu}\left[1+C\ln \left(1/t\right)\right]^{\tilde\nu},
\label{xiDS}
\end{equation}
for the infinite system, where $\nu=1$, $C$ is a disorder-dependent positive 
constant, and $\tilde\nu =1/2$; 
for $C=0$ one recovers pure-system behaviour.
As discussed in Ref.\ \onlinecite{sbl}, logarithmic corrections do not 
show up in the correlation length for finite systems, but in its
temperature derivatives;
at criticality, i.e. $t(p)=0$, Eq.\ (\ref{mul}) becomes
\begin{equation}
 {\mu_L\over L^2} \sim \left(1 - A\ \ln L\right)^{\tilde\nu},\\
\label{eq:mul2}
\end{equation}
where $A$ is some disorder-dependent constant. 
While in the Dotsenko-Shalaev theory~\cite{dstheory}, $\tilde\nu$ was
predicted
to be 1/2, here we allow it to be determined from an analysis of the data:
it is chosen in such a way that, for fixed $p$, data for 
$\left[\mu_L/L^2\right]^{1/\tilde\nu}$ versus $\ln L$ lie on a straight line,
for the largest system sizes.
Figure \ref{fig:mul} shows our results for $p=0.95$ and 0.92: we see that 
an attempt to fit our data to these logarithmic corrections would require an
unlikely variation from $\tilde\nu=0.75$ for $p=0.95$ to $\tilde\nu=100$
for $p=0.92.$
Attempts to fit the data to other forms, such as powers of $\ln L$, turned
out to be equally unsuccessful.
The conclusion is that logarithmic enhancements play no role in the bulk 
correlation length  for frustrated disorder (at least in a simple,
clearly-defined way as in the unfrustrated case).

\begin{figure}
\epsfxsize=8,5cm
\begin{center}
\leavevmode
\epsffile{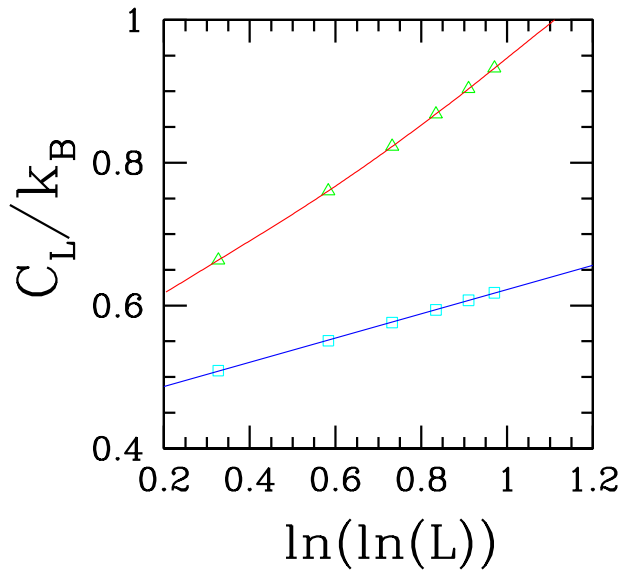}
\caption{Specific heat as a function of $\ln \ln L$, for $p=0.95$ (triangles) 
and 0.92 (squares). Error bars are smaller than data points.}
\label{fig:sph}
\end{center}
\end{figure}

Turning now to the specific heat behaviour, we recall that in
the Dotsenko-Shalaev theory, the singular part of the bulk specific
heat per particle for
the disordered Ising model, near the critical point, is  given by
\begin{equation}
C_{\infty}(t) \simeq (1/ C_0) \ln \left( 1 + C_0 \ln (1/t)\right),
\label{eq:bsph}
\end{equation}
where $C_0$ is proportional to the strength of disorder, and the pure-system 
simple logarithmic divergence is recovered as 
$C_0 \to 0$. 
For $C_0 \neq 0$ and $t \ll 1$ a double-logarithmic singularity arises, whose 
amplitude  Eq.\ (\ref{eq:bsph}) predicts to decrease as disorder increases. 
For a finite system, the usual FSS theory applied to this case 
yields\cite{dstheory} 
\begin{equation}
C_L(t=0) \simeq C_1 + a \ln \left(1 +b \ln L\right),
\label{eq:fsssph}
\end{equation}
where, similarly to Eq.\ (\ref{eq:bsph}), $b \to 0$ for vanishing disorder.
We tried to check whether such forms had any relevance in the present
case. Our results are displayed in Figure \ref{fig:sph}, and a trend
similar to unfrustrated randomness is observed: 
for low disorder, the specific heat increases with system size faster than in a 
double-logarithmic fashion (e.g., with $\ln L$);
as disorder increases ($p=0.92$), the best fit of the data crosses over to 
double-logarithmic  behaviour.
Though this may be interpreted as signalling the existence of
logarithmic corrections, such a discussion is rather subtle~\cite{msbl}.
At any rate, an inequivocal conclusion to be drawn
from our data is that the specific heat
diverges as $L\to\infty.$
Accordingly, this enables us to set $\alpha \geq 0$ in the hyperscaling
relation
$d\nu=2-\alpha$, to obtain the condition $\nu\leq 1$. 
This condition, together with the absence of logarithmic corrections for 
$\xi$, and the downturn in the sequence of estimates for $\nu$, 
lead to a scenario of $\nu(p)=1,$ as in the pure case.

\begin{figure}
\epsfxsize=8,5cm
\begin{center}
\leavevmode
\epsffile{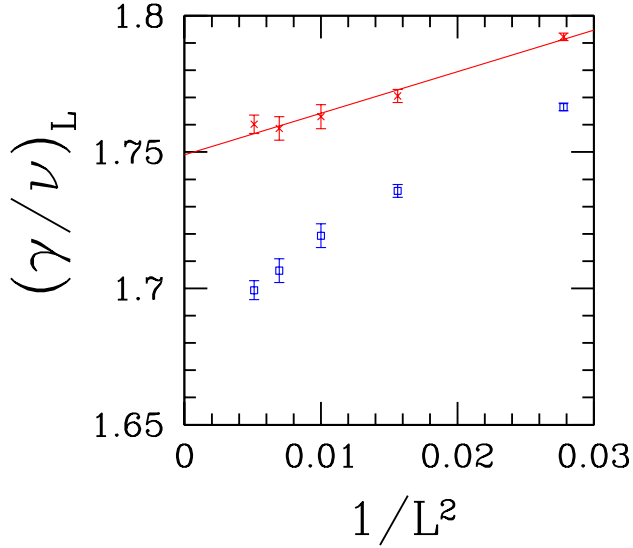}
\caption{Ratio of critical exponents, $\gamma/\nu$, as a function of $1/L^2$, 
for $p=0.95$, from Eq.\ (\protect{\ref{eq:gnu}}), calculated at $T_c$ 
determined through the scaling of $\xi^{av}$ (crosses) and of $\xi^{typ}$ 
(squares).}
\label{fig:chi}
\end{center}
\end{figure}

In order to build up a fuller picture of the low-frustration regime, we turned
to an alternate quantity, the susceptibility. 
The ratio $\gamma/\nu$ can be obtained in the usual way,
\begin{equation}
\left({\gamma\over\nu}\right)_L = 
{\ln \left[\chi_L/\chi_{L-2}\right]_{T_c}\over \ln[L/L-2]},
\label{eq:gnu}
\end{equation}
where $T_c$ is understood to be the extrapolated value.
We checked for self-consistency of critical-point
locations and properties in the following way. 
Firstly, the procedure we used above to obtain $T_c(p)$ from 
extrapolations of $\xi^{av}_L$ can be repeated for $\xi^{typ}$; 
this yields a slightly different extrapolated value,
$T_c(\xi^{typ}).$
For $p=0.95,$ for instance, one has $K_c(\xi^{typ})=0.531\pm 0.001,$
while  $K_c(\xi^{av})=0.534\pm 0.001.$
The sequence of susceptibilities calculated at these estimates of $K_c$
gives rise, through Eq.\ (\ref{eq:gnu}), to the data shown in 
Figure~\ref{fig:chi}. 
One clearly sees that the Ising value  $\gamma/\nu = 7/4$ is compatible with 
extrapolation of data calculated at $T_c(\xi^{av})$, and {\em not} with those 
coming from $T_c(\xi^{typ})$. 
We take this to mean that: 
$(i)$ from susceptibility, specific-heat and correlation-length data
the most likely self-consistent picture is one in which
the critical behaviour is pure-Ising for low frustration;
$(ii)$ though very likely $\xi^{av}$ and $\xi^{typ}$ will 
eventually scale in a similar way, higher-order corrections still produce
sizeable distortions in the accessible range of strip widths;
$(iii)$ further, it seems
that, for not very large widths, $\xi^{av}$ behaves more reliably.
 
The above analysis, together with the scaling law $\eta=2-(\gamma/\nu)$, 
predicts that, for low disorder, the exponent describing the decay of 
correlations at criticality sticks to the pure system value, $\eta={1\over 4}$.
Thus, if the exponent-amplitude relationship of conformal invariance
remains valid in the case of frustrated disorder, we should have 
$\lim_{L\to\infty}\xi_L[T_c(p)]/L=1/\pi\eta=1.273\dots$.
We obtain an estimate of $\lim_{L\to\infty}\xi_L[T_c(p)]/L$ by observing 
the trend followed by the sequences of points calculated at $K_c(p)$ and
at $K_c(p)\pm 0.001$, to determine the central estimate and its error bars
(see Fig.\ \ref{fig:xi/L});
the outcome is shown in the last column of Table I. 
In spite of the arbitrariness of this approach, the error bars thus obtained
are certainly overestimated.
Nonetheless, even with such generous allowances, the data for $p=0.92$ show 
that the conformal invariance prediction is definitely not satisfied, since 
it lies way outside the range of the error bars. 
As the critical behaviour should be the same along the critical line
(at least within the low disorder region), we are led to conclude that, 
unlike the case of unfrustrated disorder, the exponent-amplitude 
relationship of conformal invariance breaks down in the case of frustrated 
disorder.

At $p \simeq 0.89$, the transition vanishes abruptly, meaning that we do not 
find any temperature at which correlation lengths scale linearly 
with strip width.
In Ref.~\onlinecite{ozni}, it is found that the typical
correlation lengths $\xi^{typ}$
still scale linearly with $L$ at suitably low $T$ 
for a broader range of $p$--variation, along a line  that significantly
departs from the vertical on a $p-T$ diagram;
however, they find a maximum in $\xi^{typ}$ as a function of temperature for
finite values of $T$. This unexpected behaviour has indeed been observed in
studies of $\xi^{typ}$ for (unfrustrated) disordered and random-field Ising
systems~\cite{unpub}. No similar peak structure occurs when we investigate 
{\it average} correlation lengths; instead, these vary monotonically and
diverge only as $T \to 0$, consistent with the fact that a strip is essentially
one-dimensional.
Though the authors of Ref.~\onlinecite{ozni} acknowledge that such
maxima at finite $T$ are unphysical,
they assume that their data still are reliable above the peak temperatures, and
interpret the corresponding part of their low-temperature results
as marking the boundary between
a random-antiphase state and the paramagnetic regime, extending as far as
$p \simeq 0.8$ . We have not found any evidence for this phase from our
treatment. Strictly speaking, this means only that
the expected signature of the corresponding second-order phase transition
does not show up when averaged correlation lengths are considered. At present
we are unaware of why it should be so, and whether it means that 
the random-antiphase state is not present at all.  
\begin{figure}
\epsfxsize=8,5cm
\begin{center}
\leavevmode
\epsffile{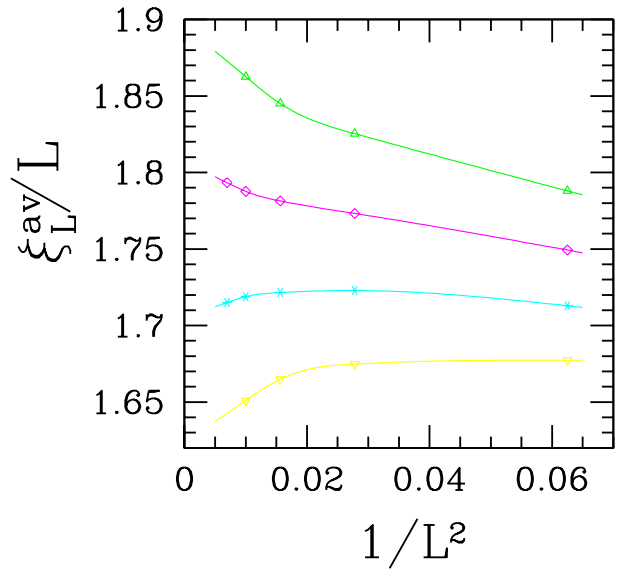}
\caption{Size-scaled averaged correlation length versus $1/L^2$ along the 
Nishimori line: 
$p=0.892$ (uptriangles), 0.891 (diamonds), 0.890 (stars), and 0.889 
(down-triangles).
For each $p$, the corresponding temperature is given by 
Eq.\ (\protect{\ref{nisheq}}). 
Error bars are smaller than data points.}
\label{fig:xiLNP}
\end{center}
\end{figure}

\section{At and below the Nishimori point}
\label{NP}

The Nishimori line is given by~\cite{nish}
\begin{equation}
\exp (2J_0/T) = p/(1-p),
\label{nisheq}
\end{equation} 
and our first task is to determine its intersect with the critical curve, 
which is done as follows. 
For each $T$, we extract $p_{\rm trial}$ from Eq.\ (\ref{nisheq}), and 
calculate $\xi^{av}(p_{\rm trial})$; 
this procedure is repeated for different system sizes, so that a sequence of 
estimates, $\xi_L/L$, is produced. 
Figure \ref{fig:xiLNP} shows our data thus obtained, and two different 
trends can be clearly observed: curves for $p=0.892$ and 0.891 display an
upward curvature, whereas those for $p=0.890$ and 0.889 are bent downwards.
Assuming a monotonic behaviour (as $L\to\infty$) of $\xi/L$, any curve outside
the interval [0.8900, 0.8910] will certainly not stabilise to a constant
value for larger $L$. 
Our central estimate for the NP is therefore just the midpoint along the 
confidence interval (or, one might say, along the complementary of the 
non-confidence domain): 
\begin{equation}
p_{N}=0.8905\pm0.0005,\ \ T_{N}=0.954\pm 0.002,
\label{ourNP}
\end{equation}
where $T_{N}$ follows from Eq.\ (\ref{nisheq}).
Our estimate for the location of the NP should be compared with those
coming from:
series work on the NL~\cite{adler},
giving $p_N= 0.886 \pm 0.003$, $T_N =0.975 \pm 0.006$;
zero-temperature calculations, together with a no-reentrance assumption,
giving  $p_N= 0.896 \pm 0.001$ or $p_N= 0.894 \pm 0.002$ (depending on
details of the fit)~\cite{kari}; exact combinatorial work~\cite{bgp}
$p_N \simeq 0.885$ [error bar presumably $\simeq 0.005$ (our estimate)];
and Monte Carlo analysis of non-equilibrium relaxation,\cite{Ozeki98}
$p_N=0.8872\pm0.0008$.
\begin{figure}
\epsfxsize=8,5cm
\begin{center}
\leavevmode
\epsffile{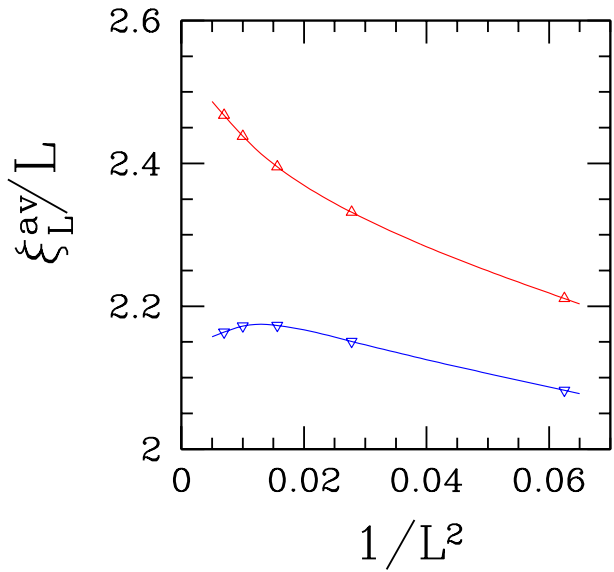}
\caption{Size-scaled averaged correlation length versus $1/L^2$ below the 
Nishimori line: 
$p=0.892$ (uptriangles) and 0.889 (down-triangles) at $T=0.4$. 
Error bars are smaller than, or of similar size to, data points.}
\label{fig:bnp}
\end{center}
\end{figure}

Once the Nishimori point has been accurately determined, we can check the
region below for reentrant behaviour. 
Evidence has been presented very recently (for Potts spin-glasses on 
hierarchical lattices) that there appears to be no fundamental reason why 
reentrances should be ruled out in thermodynamic systems~\cite{gs98}, thus
this is a matter worthy of consideration.
We examine the size dependence of $\xi^{av}$ at a temperature $T=0.4<T_N/2$ 
and at concentrations slightly away from $p_N$, $p=0.889$ and 0.892; 
the results are displayed in Fig.\ \ref{fig:bnp}.
The curve corresponding to $p>p_N$ seems to diverge as $L\to\infty$, indicating 
that the ($T,p$) point lies within the ferromagnetic region;
the one corresponding to $p<p_N$ seems to vanish as $L\to\infty$, indicating 
that the corresponding ($T,p$) point lies outside the ordered phase.
Given that both values of $p$ are very close to $p_N$, we interpret this as
an indication of absence of reentrant behaviour, and that the critical 
line $T_c(p)$ is vertical at and below the Nishimori point.
This is consistent with theoretical considerations~\cite{nish,nish2} and
with extensive numerical work~\cite{ozni,kiog,kari} specifically aimed
at the two-dimensional $\pm J$ Ising model.
As a result, we assume that the scaling directions at the NP
are, respectively, tangent to the critical curve (thus, 
purely temperature-like) and along the Nishimori
line~\cite{ldh,adler,cho}.

In order to discuss critical exponents, we note that
the numerical evaluation of temperature derivatives in Eq.~(\ref{eq:chiC}) 
implies that $\delta T \lesssim 0.001$ at the NP; since this is of the same 
order as the estimated error bars in $T_N$, we shall sit at our own central 
estimates, Eq.\ (\ref{ourNP}), and measure the corresponding $\delta T$ from 
there.

While for scaling along the tangent ({\it i.e.}, pure temperature-like) axis,
the considerations on the need to subtract free energies
of the same bond geometry~\cite{msbl} are identical to those quoted above, a
subtlety arises when considering variations along the Nishimori line, where
a temperature change implies a change in $p$ as well.
From Eq.~(\ref{nisheq}), one has $(dp/dT)_{p_N,T_N} \simeq 0.21$.
For the free energy calculation on what is supposed to be a given sample,
the use of the same pseudo-random number sequence for $T$ and $T \pm \delta T$
with the typical $\delta T = 0.007$ (to be explained 
below) means that roughly fourteen bonds in 10,000 will 
reverse sign.
We have assumed that this is the meaning of ``using the same sample'' along
the Nishimori line. While in principle the bond reversals tend to increase
in-sample fluctuations, results are manageable (albeit with relative 
error bars
$\sim$ three orders of magnitude larger than those for derivatives
along the pure--$T$ direction),
no doubt owing partly to $dp/dT$ being small at the Nishimori point.
We used a relatively large $\delta T$, compared with scaling along the
pure-$T$ axis, because of the need to compromise between {\em
fluctuations} coming from the in-sample analysis described above,
and the corresponding actual {\em variations} of the free energy, used
to approximate the derivative.
 
\begin{figure}
\epsfxsize=8,5cm
\begin{center}
\leavevmode
\epsffile{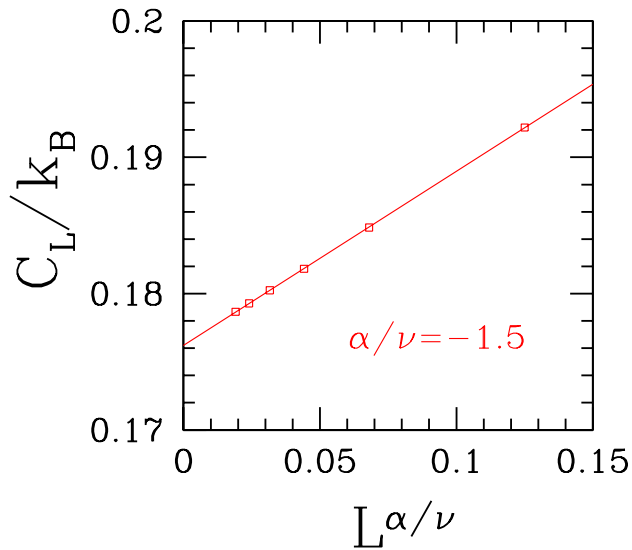}
\caption{Size dependence of the specific heat at the Nishimori point, 
resulting from the best fit of $\alpha/\nu=-1.5$ in  
Eq.\ (\protect{\ref{eq:c1}}). 
The intersect with the vertical axis is at a finite value, 0.1762.
Error bars are smaller than data points.}
\label{fig:c1}
\end{center}
\end{figure}

We have tentatively interpreted the derivatives along the NL as
specific-heat--like. Accordingly, we have applied FSS to
the finite-size specific heats along both scaling directions in order to
find estimates of $(\alpha/\nu)$; in both cases the specific heat
clearly does not diverge as $L \to \infty$.
Tangent to the boundary line we have found that attempts to fit our data to the
form 
\begin{equation}
C_L = C_{\infty} +a L^{\left({\alpha\over\nu}\right)_{\rm trial}},
\label{eq:c1}
\end{equation}
yield a much smaller (four orders of magnitude)
chi-square for $(\alpha/\nu)_{trial} = -1.5$ than for 
$(\alpha/\nu)_2 \simeq -1.1$ 
(the latter is extracted from $\nu \simeq 2.2$ of Ref.~\onlinecite{cho} plus 
the hyperscaling relation $d\nu = 2 - \alpha$); 
Figure~\ref{fig:c1} shows that our data fit neatly into a single-power
form, {\it i.e.}, corrections to scaling seem of little relevance in the
case.
Along the Nishimori line our data do not give a satisfactory behaviour
of chi-square for any sensible fitting to Eq.\ (\ref{eq:c1}):
varying $\alpha/\nu$ between $-2$ and $-0.5$ does not change chi-square 
significantly, and this persists even when corrections to scaling are 
accounted for. 
Thus we are not in a position to compare these data to the percolation 
value~\cite{stauffer}  $(\alpha/\nu)_p = -1/2$.
In  Table~\ref{table2}
 we display our raw data, so readers can reproduce the analysis
quoted above, and try alternative procedures of their own devising.   

\begin{table}
\caption{Second derivatives of free energy at the NP, for strip widths 
$L=4-14$. $T$: along the temperature axis; $NL$: along the Nisihmori
line (see text). Uncertainties in last digits are given in parenthesis. }
\vskip 0.2cm 
 \halign to \hsize{\hskip 1.6cm\hfil#\hfil&\hfil#\hfil&\hfil#\hfil\cr
  $L$ & $T$       & $NL$\cr
4  & \ 0.192183(7)  & \ 1.558(38) \cr
6  & \ 0.184854(10) & \ 1.704(44) \cr
8  & \ 0.181824(13) & \ 1.772(39) \cr
10 & \ 0.180245(7)  & \ 1.846(30) \cr
12 & \ 0.179288(25) & \ 1.846(67) \cr
14 & \ 0.178667(20) & \ 1.859(73) \cr}
\label{table2}
\end{table}

\begin{figure}
\epsfxsize=8,5cm
\begin{center}
\leavevmode
\epsffile{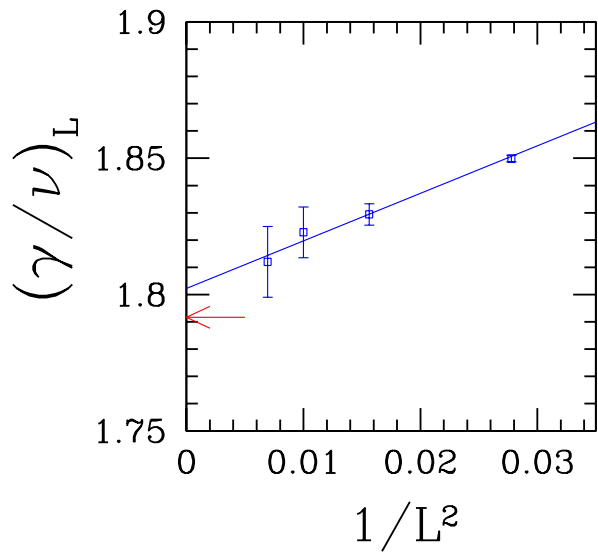}
\caption{Ratio of critical exponents, $\gamma/\nu$, as a function of $1/L^2$, 
at the Nishimori point, from Eq.\ (\protect{\ref{eq:gnu}}). Arrow points
to percolation value (see text).}
\label{fig:gnuNP}
\end{center}
\end{figure}

We have thus turned to calculating the uniform susceptibility;
as it couples to a ferromagnetic order parameter, the corresponding value 
of $\gamma/\nu$ is related to criticality upon crossing of
the ferro-paramagnetic boundary ({\it i.e.} along the Nishimori line).
In Fig.\ \ref{fig:gnuNP} we show $\left(\gamma/\nu\right)_L$, calculated
from Eq.\ (\ref{eq:gnu}), with $T_c\equiv T_N$, as a function of $1/L^2$. 
The extrapolated value $1.80 \pm 0.02$ (where the estimated error bars
are subjective, but certainly conservative) 
compares favourably with 
$(\gamma/\nu)_p = 43/24=1.7917...$ of percolation.
Series work\cite{adler} gives $\gamma=2.37\pm0.05$ and $\nu=1.32\pm0.08$,
which yield $\gamma/\nu=1.80\pm0.09$.

Finally, correlation lengths obtained from the decay of spin-spin
correlations (which therefore couple to a ferromagnetic order parameter)
give an extrapolation of $\xi^{av}/L$ to $1.75\pm 0.05$ at the NP
(see Fig.\ \ref{fig:xiLNP}),
rather different from the percolation $(\pi \eta_p)^{-1} = 1.5279 \dots$
From experience for low frustration, as described above, we interpret
this as signalling a breakdown of the exponent-amplitude relationship,
rather than indicating that the transition is not in the percolation 
universality class.

\section{CONCLUSIONS}
\label{concl}

We have studied the asymmetric $\pm J$ Ising model on a square lattice,
by means of transfer matrix calculations of several quantities on
long, finite-width strips.
Firstly, use has been made of a configurationally {\em averaged} correlation 
length, which is distinct from the {\em typical} (or most probable) 
correlation length, used in previous transfer matrix studies of the same model.
We have shown that an intrinsically self-consistent picture can be
obtained by the use of the former quantity, while (at least
for the strip widths within reach) it seems that higher-order
corrections to scaling may distort analyses based on the latter.
Points on the critical curve $T_c(p)$ were then obtained as those for which 
$\xi^{av}/L$ approached a finite value as $L\to\infty$;
to the best of our knowledge, the estimates for $T_c$ at $p=0.99,$ 0.95, and
0.92 thus obtained are the most accurate to date.
Secondly, the critical behaviour along the critical line has been discussed 
through the analysis of $d\xi^{av}/dT$, as well as in terms of other 
quantities, such as the specific heat and the zero-field susceptibility.

The following picture emerged from our analysis.
Above the Nishimori line, the correlation length and the susceptibility appear
to diverge with power laws, with the same exponents as in the pure case, 
$\nu=1$ and $\gamma=7/4;$
logarithmic corrections (i.e., enhancements) do not seem to play a
role in the behaviour of these quantities. 
We were also able to establish that the specific heat diverges, though at most 
logarithmically.
Further, the exponent-amplitude relationship of conformal invariance 
breaks down as a result of frustration.
These results are in marked contrast with the case of unfrustrated disorder, 
for which logarithmic enhancements were needed in order to explain an apparent 
disorder dependence on estimates for $\nu$, and the conformal invariance 
prediction applies.

The intersection of the Nishimori line (NL) with the critical curve ($T_N,p_N$)
has been determined, near which the critical behaviour was analysed;
our estimates for ($T_N,p_N$) are also the most accurate to date.
Approaching this Nishimori point either along a temperature-like direction or
along the NL, one finds non-diverging specific heats; 
while for the former we were able to extract 
$\left(\alpha/\nu\right)_T\simeq -1.5$, for the latter we could not find 
reliable fits.
However, analysis of the uniform susceptibility,
which probes the phase transition along the Nishimori line, 
showed  percolation-like behaviour,
in the sense that $\gamma/\nu$ is very close to the percolation value.
Conformal invariance is also absent at the Nishimori point.
Further work is clearly necessary in order to fully elucidate all the 
subtleties related to the multicritical behaviour at the Nishimori point.

Finally, we found no signature of a reentrance in the phase diagram 
below the Nishimori point;
instead, the critical curve below this point seems to be parallel to the
temperature direction.

\acknowledgements 
We thank Laborat\'orio Nacional de Computa\c c\~ao Cien\-t\'\i\-fica 
(LNCC) for use of their computational facilities, and
Brazilian agencies  CNPq and FINEP, for financial support.
SLAdQ thanks the Department of Theoretical Physics
at Oxford, where part of this work was done, for the hospitality, and
the cooperation agreement between CNPq and
the Royal Society for funding his visit. Special thanks are due to R. B.
Stinchcombe for invaluable discussions, and to D. Stauffer for useful
suggestions.


\begin{references}
\bibitem{rieger} H. Rieger in {\it Annual Reviews of Computational
Physics},
 edited by D. Stauffer (World Scientific, Singapore, 1995) Vol. 2.
\bibitem{ea} S. F. Edwards and P. W. Anderson, J. Phys. F {\bf 5}, 965 (1975).
\bibitem{byom} R. N. Bhatt and A. P. Young, Phys. Rev. Lett. {\bf 54}, 924 
(1985); A. T. Ogielski and I. Morgenstern, {\it ibid. \bf 54}, 928 (1985).
\bibitem{by} K. Binder and A. P . Young, Rev. Mod. Phys. {\bf 58}, 801 
(1986).
\bibitem{lc} N. Lemke and I. A. Campbell, Phys. Rev. Lett. {\bf 76}, 4616 
(1996).
\bibitem{parisi} G. Parisi, J. J. Ruiz-Lorenzo and D. A. Stariolo,
J. Phys. A {\bf 31}, 4657 (1998).
\bibitem{domany} E. Domany,  J. Phys. C {\bf 12}, L119 (1979).
\bibitem{jamet} J. P. Jamet, C. P. Landee, J. Ferr\'e,
M. Ayadi, H. Gaubi and I. Yamada, J. Phys.: Cond. Matt. {\bf 8},
5501 (1996).
\bibitem{schins} A. G. Schins, M. Nielsen, A. F. M. Arts and H. W.
de Wijn, \prb {\bf 49}, 8911 (1994); J. Magn. Mag. Mater. {\bf 140-144},
1715 (1995); A. P. Ramirez, A. G. Schins, A. F. M. Arts and H. W. de
Wijn, J. Magn. Mag. Mater. {\bf 140-144}, 1713 (1995).
\bibitem{ozni} Y. Ozeki and H. Nishimori, J. Phys. Soc. Japan {\bf 56}, 3265
(1987).
\bibitem{kiog} H. Kitatani and T. Oguchi, J. Phys. Soc. Japan {\bf 61}, 1598
(1992).
\bibitem{nish} H. Nishimori, Prog. Theo. Phys. {\bf 66}, 1169 (1981).
\bibitem{ldh} P. Le Doussal and A. B. Harris, Phys.Rev. Lett. {\bf 61}, 625 
(1988).
\bibitem{adler} R. R. P. Singh and J. Adler, Phys. Rev. B {\bf 54}, 364 (1996).
\bibitem{cc} J. T. Chalker and P. D. Coddington, J. Phys. C {\bf 21}, 2665 
(1988).
\bibitem{cho} S. Cho and M. P. A. Fisher, Phys. Rev. B {\bf 55}, 1025 (1997).

\bibitem{stauffer} 
 D. Stauffer and A. Aharony, {\it Introduction to Percolation Theory}
(Taylor and Francis, London, 1991).

\bibitem{weak}
J.-K. Kim and A. Patrascioiu, \prl {\bf 72}, 2785 (1994);
R. K\"uhn,  \prl {\bf 73}, 2268 (1994); 
M.\ F\"ahnle, T.\ Holey, and J.\ Eckert,
J.\ Mag.\ Magn.\ Mat.\ {\bf 104--107}, 195 (1992).

\bibitem{dstheory}
Vik.S. Dotsenko and Vl.S. Dotsenko, J. Phys. C {\bf 15}, 495 (1982); 
B. N. Shalaev, Phys. Rep. {\bf 237}, 129 (1994);
W. Selke, L. N. Shchur and A. L. Talapov 
in {\it Annual Reviews of Computational Physics} Vol. 1, edited by 
D. Stauffer (World Scientific, Singapore, 1994).

\bibitem{cardy}
J.L. Cardy, in {\it Phase Transitions and Critical Phenomena},
edited by C. Domb and J.L. Lebowitz (Academic, New York, 1987), vol. 11.

\bibitem{sbl} F. D. A. Aar\~ao Reis, S.L.A. de Queiroz and R. R. dos Santos, 
Phys. Rev. B {\bf 54}, R9616 (1996); {\bf 56}, 6013 (1997).
\bibitem{msbl} D. Stauffer, F. D. A. Aar\~ao Reis, S.L.A. de Queiroz
 and R. R. dos Santos, Int. J. Mod. Phys. C {\bf 8}, 1209 (1997).

\bibitem{sldq}
S.L.A. de Queiroz, Phys. Rev. E {\bf 51}, 1030 (1995).
\bibitem{cj} J. L. Cardy and J. L. Jacobsen, \prl {\bf 79}, 4063 (1997);
J. L. Jacobsen and J. L. Cardy, Nucl. Phys. {\bf B515}, 701 (1998).
\bibitem{berche} C. Chatelain and B. Berche, \pre {\bf 58}, R6899(1998).
\bibitem{nig82} 
M.~P.~Nightingale, J.~Appl.~Phys. {\bf 53}, 7927 (1982).
\bibitem{fs1}
M.N. Barber, in {\it Phase Transitions and Critical Phenomena}, 
edited by C. Domb and J.L. Lebowitz (Academic, New York, 1983), Vol. 8.
\bibitem{fs2}
M.P. Nightingale, in {\it Finite Size Scaling and Numerical Simulations
of Statistical Systems}, edited by V. Privman (World Scientific, Singapore, 
1990).
\bibitem{dqrbs}
S.L.A. de Queiroz and R. B. Stinchcombe, Phys. Rev. E {\bf 54}, 190 (1996).
\bibitem{dds} 
B. Derrida and L. de Seze, J. Phys. (France) {\bf 43}, 475 (1982).
\bibitem{glaus}
U.Glaus, J. Phys. A {\bf 20}, L595 (1987). 
\bibitem{ranmat} A. Crisanti, G. Paladin and A. Vulpiani, 
{\it Products of
Random Matrices in Statistical Physics}, Springer Series in Solid State Sciences
 Vol. 104, edited by Helmut K. Lotsch (Springer, Berlin, 1993).
\bibitem{derv} B. Derrida and J. Vannimenus, Phys. Rev. B {\bf 27}, 4401 (1983).
\bibitem{lyapcor} S.L. A. de Queiroz, J. Phys. A {\bf 30}, L443 (1997).
\bibitem{unpub} S. L. A. de Queiroz and R. B. Stinchcombe (unpublished).
\bibitem{gs98} M. J. P. Gingras and E. S. S\o rensen, \prb {\bf 57}, 10\, 264
(1998).
\bibitem{nish2} H. Nishimori, J. Phys. Soc. Japan {\bf 55}, 3305 (1986).
\bibitem{kari} N. Kawashima and H. Rieger, Europhys. Lett {\bf 39}, 85 (1997).
\bibitem{bgp} J. A. Blackman, J. R. Gon\c calves, and J. Poulter,
\pre {\bf 58}, 1502 (1998).
\bibitem{Ozeki98} Y.\ Ozeki and N.\ Ito, J.\ Phys.\ A {\bf 31}, 5451 (1998).
\end{references}
\end{document}